\documentclass[11pt,twoside]{article}
\usepackage{asp2010}

\resetcounters

\begin{document}

\title{DAS: a data management system for instrument tests and operations} 
\author{Marco~Frailis$^1$, Stefano~Sartor$^1$, Andrea~Zacchei$^1$,
  Marcello~Lodi$^3$, Roberto~Cirami$^1$, Fabio~Pasian$^1$,
  Massimo~Trifoglio$^2$, Andrea~Bulgarelli$^2$, Fulvio~Gianotti$^2$,
  Enrico~Franceschi$^2$, Luciano~Nicastro$^2$, Vito~Conforti$^2$,
  Andrea~Zoli$^2$, Ricky~Smart$^4$, Roberto~Morbidelli$^4$, and
  Mauro~Dadina$^2$
\affil{$^1$INAF-Osservatorio Astronomico di Trieste, Via G.B. Tiepolo 11, Trieste, Italy}
\affil{$^2$INAF-IASF Bologna, Via P. Gobetti 101, Bologna, I40129, Italy}
\affil{$^3$Telescopio Nazionale Galileo, FGG - INAF, Rambla Jos\'e Ana F\'ernandez P\'erez 7, 38712 Bre\~na baja, Espa\~na}
\affil{$^4$INAF-Osservatorio Astrofisico di Torino, Strada Osservatorio 20, 10025 Pino Torinese, Italy}}

\begin{abstract}
  The Data Access System (DAS) is a metadata and data management
  software system, providing a reusable solution for the storage of
  data acquired both from telescopes and auxiliary data sources during
  the instrument development phases and operations. It is part of the
  Customizable Instrument WorkStation system (CIWS-FW), a framework
  for the storage, processing and quick-look at the data acquired from
  scientific instruments. The DAS provides a data access layer mainly
  targeted to software applications: quick-look displays,
  pre-processing pipelines and scientific workflows. It is logically
  organized in three main components: an intuitive and compact Data
  Definition Language (DAS DDL) in XML format, aimed for user-defined
  data types; an Application Programming Interface (DAS API),
  automatically adding classes and methods supporting the DDL data
  types, and providing an object-oriented query language; a data
  management component, which maps the metadata of the DDL data types
  in a relational Data Base Management System (DBMS), and stores the
  data in a shared (network) file system. With the DAS DDL, developers
  define the data model for a particular project, specifying for each
  data type the metadata attributes, the data format and layout (if
  applicable), and named references to related or aggregated data
  types. Together with the DDL user-defined data types, the DAS API
  acts as the only interface to store, query and retrieve the metadata
  and data in the DAS system, providing both an abstract interface and
  a data model specific one in C, C++ and Python. The mapping of
  metadata in the back-end database is automatic and supports several
  relational DBMSs, including MySQL, Oracle and PostgreSQL.
\end{abstract}

\section{Introduction}
The Data Access System (DAS) is a reusable software system that allows
storage, retrieval and management of metadata and data acquired from
telescopes and auxiliary data sources or produced by their subsequent
levels of data processing.  It is part of the Customizable Instrument
WorkStation software project \citep[see][]{conforti_adassxxiii}, aimed
at providing a framework, named CIWS-FW, for the storage, processing
and quick-look at data acquired from space-borne and ground-based
telescope observatories, to support all their development phases. The
DAS inherits several high-level design concepts from the \textit{Planck}-LFI
Data Management Component \citep{hazell_2002}, a system designed and
developed by the MPA (Max Planck institute for Astrophysics) \textit{Planck}
Analysis Centre and successfully deployed by the \textit{Planck}-LFI (Low
Frequency Instrument) Data Processing Centre for the instrument tests
and flight operations and used by all its processing levels
\citep{zacchei_2011}.

\articlefigure[width=0.9\textwidth]{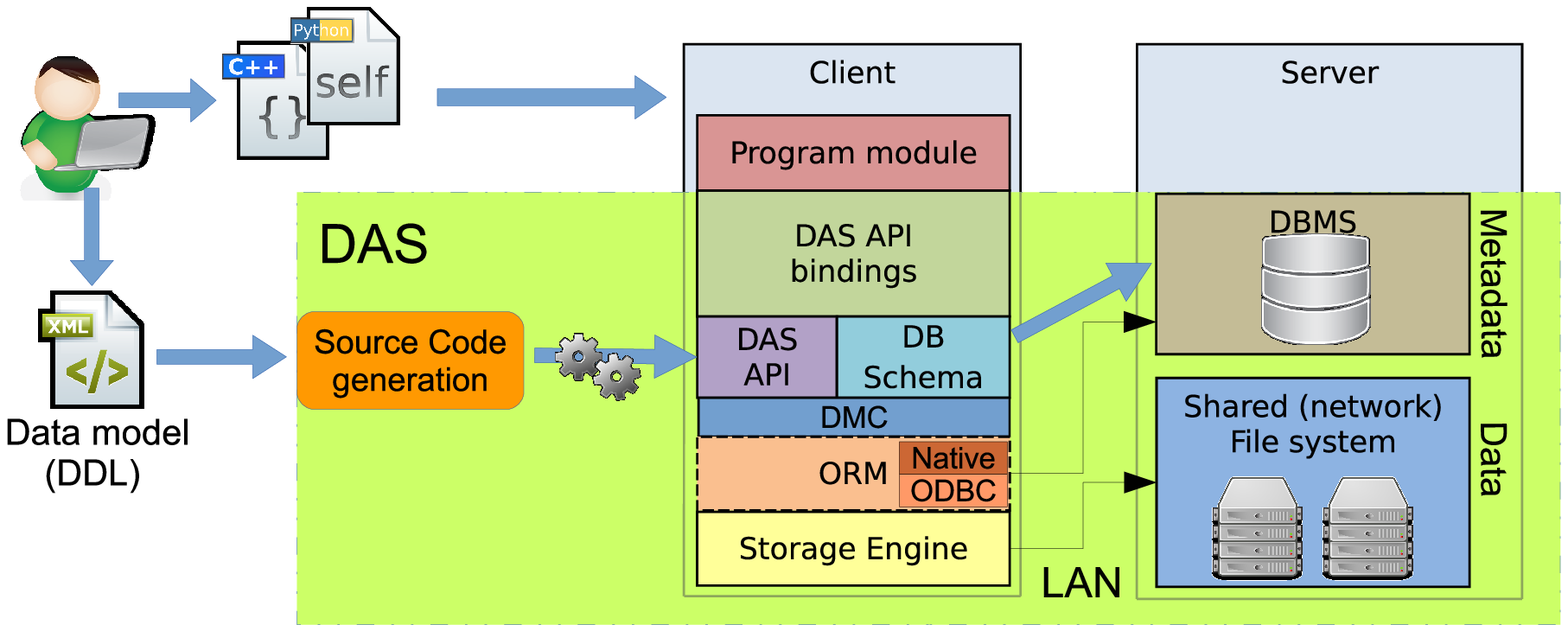}{das_system}{DAS system overview}

The DAS provides a data access layer mainly targeted to software
applications: quick-look displays for the near real-time or off-line
analysis of the instrument raw data, pre-processing and calibration
pipelines, data analysis tasks or data reduction workflows. It is
logically organized in three main components. A {\bf Data Definition
  Language} ({\bf DAS DDL}) in XML format, consisting of an intuitive
and compact grammar that the developers use to define the data
structures that will be archived and retrieved in a specific
project. Each DDL type is defined by specifying its metadata
attributes, data format and layout - binary table or image - and
associations (references) to other related types.  An {\bf Application
  Programming Interface} ({\bf DAS API}), automatically mapping each
user-defined DDL type to a number of supporting classes and methods,
and providing an object-oriented query language on the metadata
attributes and the associations of the DDL types. A {\bf Data Management
Component} ({\bf DMC}), which maps the metadata and the associations of
the DDL data types to a relational DBMS, with the aid of an
Object-to-Relational Mapping system (ORM), and stores the data in a
shared (network) file system.

The first DAS user task consists in the specification of the data
model for a particular project, by creating a DDL file that collects
all the data structure definitions. Starting from this file, the DAS
building system automatically generates classes that correspond to the
DDL data types and updates the DAS API library and the DBMS database
schema with the new types (see fig. \ref{das_system}). Then, the DAS
user can code the program module, using the DAS API to persist, query
and access the data.

\section{DAS Data Definition Language}
The DDL provides an XML grammar, formalized in the XML Schema
Definition language (XSD), to define new data types that will be
stored in the DAS system. 

\articlefigure[width=0.7\textwidth]{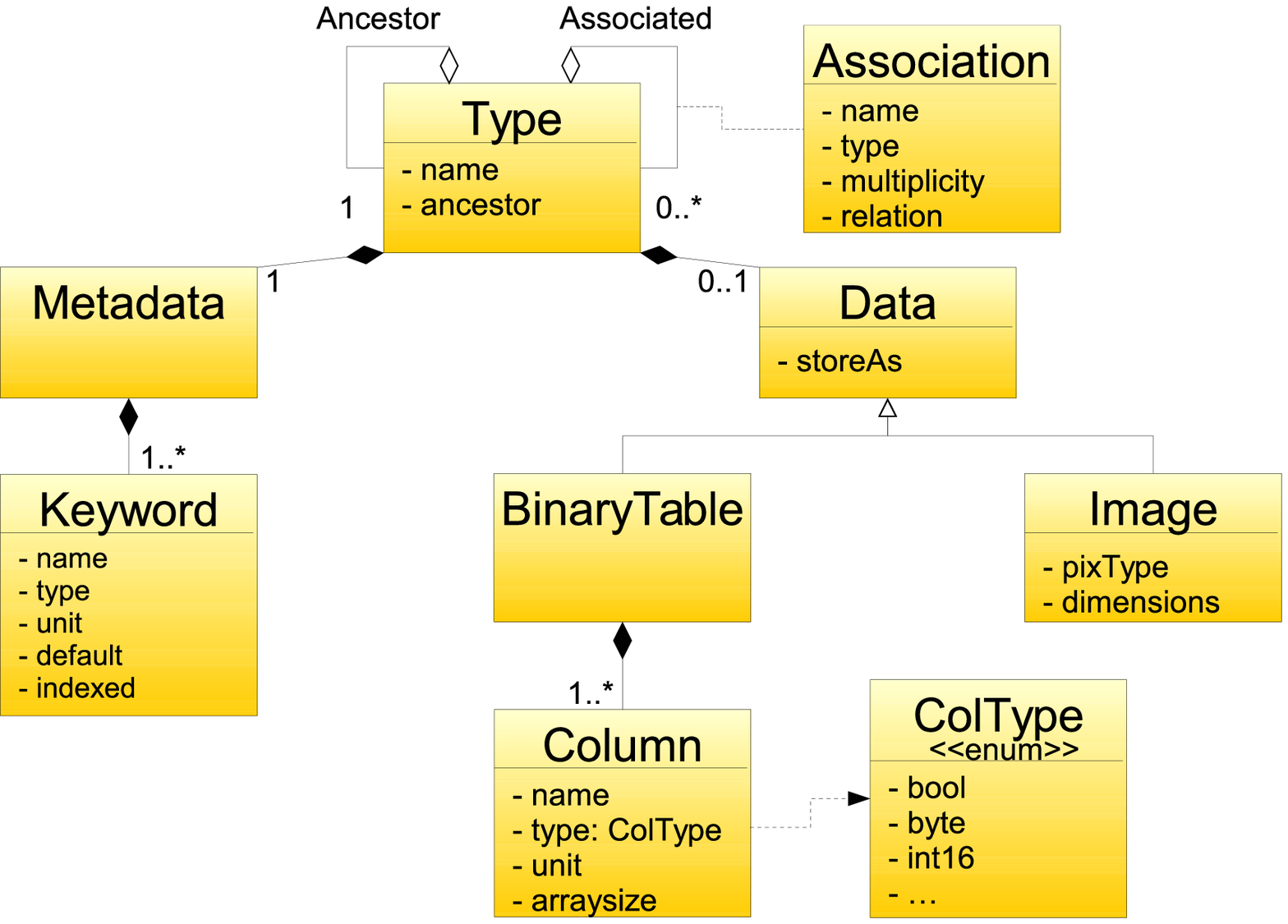}{das_ddl}{DDL grammar elements}

A DDL data type definition includes two main components: metadata and
data (see fig. \ref{das_ddl}). The metadata section defines a list of keywords describing the
data. The data section can define either a binary table or an
image. These two components are logically analogous to the header block and data
block of the FITS standard.  Since the data section is not mandatory,
it is possible to define metadata-only data types.  Additionally, a
data type can be associated to other correlated data types. A single
association with a data type can also specify: i) a multiplicity, when
many instances of the same type are involved in an association; ii)
the relation type ({\em shared}, {\em exclusive}, {\em extend}), that affects the
mapping of the association in the database schema. In order to define
new data types by extending existing ones, in a data type definition
it is possible to add a single parent type (ancestor).  

By default, the DAS system stores the metadata in a back-end
relational DBMS; the data is transparently persisted in a binary
format (analogous to the VOTable binary serialization, \citet{Ochsenbein_2011}) in a shared
file system. However, for small data sizes (small arrays or images), a
DDL type definition can specify to store the data as BLOBs (Binary
Large Objects) in the DBMS.

\section{DAS Architecture and API}
Figure \ref{das_system} shows the high-level components of the DAS
system. The core of the system is developed in C++. One component is
the ORM system, that allows to persist C++ objects to a relational
DBMS, automatically handling the conversion between C++ and SQL
types. We are using an existing open-source ORM, ODB
(www.codesynthesis.com), which currently supports several DBMSs,
including MySQL, Oracle, PostreSQL and SQLite. For each supported
DBMS, ODB uses the native C API instead of ODBC to reduce the
overhead.  

From the DDL file provided by the user, the DAS automatically
generates the C++ classes corresponding to each type and the ORM
directives. The database support code is then built through the ORM
system. The DMC implements the DAS API, using some concepts taken from
the JPA specification (Java Persistency API), in particular the so
called persistence context \citep{bauer_2006}, which simplifies the
automatic synchronization of an in-memory data object and all its
associated instances with the persistent counterparts.  The Data
Storage Engine component implements the data serialization
mechanism. Currently, only one binary serialization is provided;
however base classes are available to support other possible types of
serialization.

\articlefiguretwo{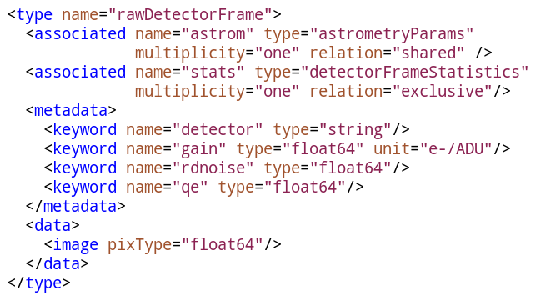}{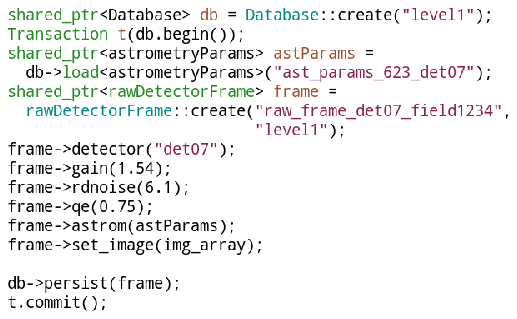}{das_example}{DDL type definition and client code example}

Figure \ref{das_example} shows a short example of DDL data type
definition in XML and the client code in C++ using the class
automatically generated for that type by the DAS system. The DAS core
API, in C++, provides both a template based and a polymorphic
interface (for run-time type inference). Currently, a Python binding,
based on SWIG, is also available.

\section{Conclusions}
The DAS requirements and subsequent design and development is based on
the experience gained by the team in monitoring, control and analysis
software for space-borne and ground observatories (\textit{Planck},
AGILE, TNG, GSCII, REM). It will soon reach a first stable release in
the upcoming months. Additional information on the CIWS-FW and the DAS
system can
be found at the following links:\\
\url{http://ciws-fw.iasfbo.inaf.it/ciws-fw}, \\
\url{http://redmine.iasfbo.inaf.it/projects/ciwsfw/repository/gitdas}.

\acknowledgements The CIWS-FW is being developed through the TECNO-INAF-2010 technological
project funded by INAF.

\bibliography{P045}

\end{document}